\documentclass[]{interact}
\usepackage{epstopdf}
\usepackage{caption}
\usepackage{subcaption}
\usepackage{natbib}
\bibpunct[, ]{(}{)}{;}{a}{}{,}
\renewcommand\bibfont{\fontsize{10}{12}\selectfont}
\makeatletter
\@ifundefined{amscode}{%
  \newenvironment{amscode}{\par\medskip\noindent{\itshape 2010 Mathematics Subject Classification}: }{\par}%
}{}
\makeatother
\theoremstyle{plain}

\theoremstyle{definition}

\theoremstyle{remark}

\begin{document}
\title{Conditional Distribution Estimation Given Functional Covariates Using Deep Operator Networks}
\author{
\name{Bingqing Hu, Ran Zou, and Bin Nan \thanks{CONTACT Bin Nan. Email: nanb@uci.edu}}
\affil{Department of Statistics, University of California, Irvine, CA 92697, USA}
}
\maketitle
\begin{abstract}
Functional linear models are commonly used for analyzing functional data with scalar responses, particularly for modeling the conditional mean of the response given functional covariates. In this work, we extend the traditional scalar-on-function regression in two major aspects: 1. estimating the conditional distribution function instead of a particular characteristic such as the conditional mean; 2. considering an arbitrary operator without imposing functional linearity or any model assumption. We use a likelihood approach for the conditional hazard function and apply convolutional neural networks to approximate the effects of functional inputs and estimate the conditional distribution using deep operator networks. Through simulations and a real world data example, we show the desirable robustness of the proposed method in comparison with the mean regression neural networks, demonstrating that our approach achieves better conditional distribution estimation and interval prediction in complex data settings.
\end{abstract}
\begin{amscode}
62G05; 62G08; 62M45
\end{amscode}
\section{Introduction}
Functional Data Analysis (FDA) \citep{ramsay, wang2016functional}  is a branch of statistics that deals with data in the form of curves or functions. Unlike traditional data, functional data are intrinsically infinite dimensional and often generated by underlying continuous-time processes. In reality, the functional data are usually collected discretely over time or space. It might be tempting to consider functional data as classical multivariate data. Unfortunately, multivariate manipulation ignores the functional features of the data, and the curse of dimensionality occurs.  Numerous practical studies have shown that a direct functional approach gives better results \citep{funcMLP}. In this work, we consider the setting where the response variable is a scalar and at least one of the predictors is a random function. Specifically, let $Y$ be a response variable and $X(t)$, $t \in \mathbb{T}$, be potentially multiple functional covariates. The conventional FDA focuses on a functional linear model (FLM) \citep{flm99,flm} that has the following form:
\begin{equation*}
    E(Y |  X) = \beta_0 + \int X(s)\beta(s)ds,
\end{equation*}
which focuses on a linear functional effect of $X(t)$ on $Y$. FLM can be generalized by using a nonlinear link function between the functional linear predictor and the response variable \citep{gflm}.
In recent years, as modern deep learning methods have shown competitive performance across various fields, researchers have started applying them to FDA. Most work uses certain basis functions to expand functional inputs then feeds them into a standard feed forward neural network. For example, \cite{funcMLP} and \cite{mlp_func_05} use functional neurons in the first layer of a neural network, where the functional input is approximated by a finite number of B-splines. \cite{dlfunctionalinput} consider expanding the functional covariate using Fourier basis functions in their neural network approach. The basis coefficients are updated as the neural network learns. \cite{fda-adaptive_basis} propose an alternative neural network architecture that consists of a novel basis layer implemented via micro-networks, where basis functions themselves do not need to be a priori and can be learned from the data.
On the other hand, convolutional neural networks (CNNs) \citep{cnn_original} are designed to handle high-dimensional inputs and naturally pool neighboring information. It also enjoys an important theoretical property that its convergence rate is independent of the input dimension  under suitable assumptions \citep{convergence_cnn_2}. CNN-based methods have achieved great success in imaging data analysis and many other fields, revolutionizing the analysis of complex datasets across various domains. Compared with feed forward neural networks using basis expansions for the mean regression in FDA, CNNs are more straightforward, allowing for automatic learning of hierarchical representations from raw data. As shown in \cite{dlfunctionalinput}, CNNs achieve comparable performance to basis expansion methods in some functional datasets.
In this work, we consider applying CNNs as a nonparametric approach to estimate the conditional distribution function of a  response variable $Y$ given functional covariates $X$ that are not necessarily smooth functions. Estimating the conditional distribution function of a continuous response variable given multiple Euclidean covariates using traditional nonparametric methods, e.g., the kernel smoothing, is a challenging statistical problem  \citep{hall-aos}. It is even more challenging when functional covariates are involved. Recently, \cite{hu&nan} have developed a nonparametric maximum likelihood approach using fully connected feed forward neural networks for the conditional distribution function estimation given multiple Euclidean covariates. We propose to extend the work of \cite{hu&nan} using the deep operator network (DeepONet) \citep{deeponet} to estimate the conditional distribution function of $Y$ given functional covariates $X$. Here we include Euclidean covariates (if there is any) in $X$ and denote $X$ as functional covariates for notational simplicity.
\section{Methodology}
\label{sec:meth}
\begin{figure}[h]
    \centering
    \includegraphics[width=0.75\textwidth]{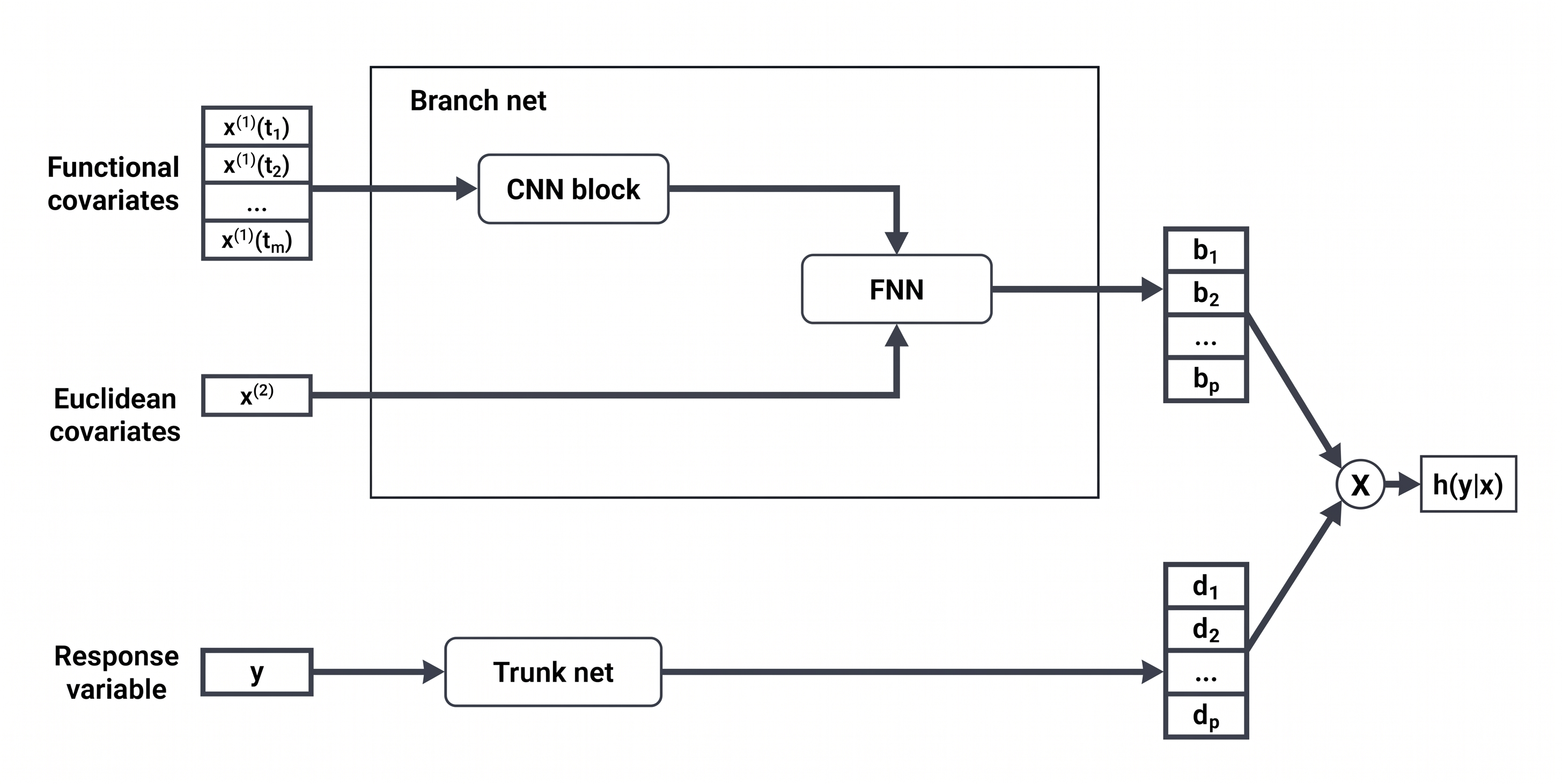}
    \caption{DeepONet structure with CNN in the branch net.}
    \label{fig:don_cnn}
\end{figure}

Let $\{(x_i(\cdot),y_i)\}_{i=1}^n$ denote a dataset of independent and identically distributed (i.i.d.) copies of functional covariates $X(\cdot)$ and the associated continuous response variable $Y$. Without loss of generality, in the following we denote $X(t)$ to be a stochastic process defined over a bounded interval $t \in [a,b]$, which is observed on a set of grid points $a\le t_1<t_2<\cdots <t_m \le b$.
We aim to estimate the conditional cumulative distribution function (CDF) $F(y  |  X)=P(Y \le y  |  X)$. Following \cite{hu&nan}, we construct a nonparametric likelihood function for the hazard function:
\[
\lambda(y  |  X) = \lim_{\Delta y \to 0} \frac{P(y \leq Y < y + \Delta y  |  Y \geq y, X)}{\Delta y}.
\]
Let $h(y  |  X) = \log \lambda(y |  X)$. Then $h$ can take any real value without any constraint. The conditional CDF can be expressed in terms of the hazard function as:
\begin{equation}
    F(y  |  X) = 1 - \exp \left( - \int_{-\infty}^y e^{h(s  |  X) \,} ds \right).
    \label{eq: cdf_functional}
\end{equation}
Once an estimator of $h$ becomes available, an estimator of the conditional CDF can be obtained from the above equation (\ref{eq: cdf_functional}) given $X$.
Clearly $h(y|X)$ is an operator for a functional covariate $X$. We propose to apply the recently developed DeepONet method \citep{deeponet} for the estimation of $h(y|X)$. A DeepONet consists of two sub-networks, one for encoding the functional input at grid points $\{t_1,t_2,...t_m\}$ (branch net), and another for encoding the location $y$ for the output function (trunk net). The branch-trunk architecture is inspired by the Universal Approximation Theorem for operators \citep{operator_universal}, which states that two fully connected neural networks with a single hidden layer, combined by a vector dot product of the outputs, are able to approximate any continuous nonlinear operator with arbitrary accuracy. Specifically, suppose that $\sigma$ is a continuous non-polynomial function, $\mathbb{T}$ is a Banach space, $K_1 \subset \mathbb{T}$ and $K_2 \subset \mathbb{R}^d$ are compact sets in $\mathbb{T}$ and $\mathbb{R}^d$, respectively, $V$ is a compact set in $C(K_1)$, and $h$ is a nonlinear continuous operator which maps $V$ into $C(K_2)$. Here $C(K)$ denotes the Banach space of all continuous functions defined on $K$ equipped with the uniform norm. For any $\epsilon > 0$, there are positive integers $n$, $p$, and $m$, constants $c_i^k$, $\xi_{ij}^k$, $\theta_i^k$, $w_k \in \mathbb{R}$, $t_j \in K_1$, $i = 1, \ldots, n$, $k = 1, \ldots, p$, and $j = 1, \ldots, m$, such that
\begin{equation} \label{eq:UAT}
\left| h(x)(y) - \sum_{k=1}^p \sum_{i=1}^n c_i^k \sigma \left( \sum_{j=1}^m \xi_{ij}^k x(t_j) + \theta_i^k \right) \sigma (w_k \cdot y + \zeta_k) \right| < \epsilon
\end{equation}
holds for all \(x \in V\) and \(y \in K_2\). Note that we use $h(x)(y)$ and $h(y|x)$ interchangeably to denote the operator of interest, and we have $K_1 = [a,b]$ and $d=1$ in our case.
A generalized version of the approximation theorem (Theorem 2 in \cite{deeponet}) states that for any nonlinear continuous operator $h$ and any $\epsilon>0$, there exist positive integers $m$, $p$, continuous vector functions $g: \mathbb{R}^m \rightarrow \mathbb{R}^p$, $f: \mathbb{R}^d \rightarrow \mathbb{R}^p$, and $t_1, t_2, \ldots, t_m$, such that,
\begin{equation}
    |h(x)(y)-\langle g[x(t_1),x(t_2),...,x(t_m)],f(y)\rangle| < \epsilon
\end{equation}
holds for all $x$ and $y$, where $\langle \cdot, \cdot \rangle$ denotes the dot product in $\mathbb{R}^p$. The functions $g$ and $f$ can be chosen as diverse classes of neural networks, for example, fully connected neural networks, residual neural networks, or convolutional neural networks.
In practice, one may consider using multiple hidden layers and different neural network architectures for both branch and trunk networks. Here we use a two-layer fully connected neural network for the trunk net. For the branch net, we first process functional covariates $x^{(1)}$ using a CNN block that consists of two 1D convolutional layers, each followed by a ReLU activation and max-pooling, then concatenate the resulting feature vector directly with Euclidean covariates $x^{(2)} \in \mathbb{R}^d$, which pass through a two-layer fully connected neural network to produce basis coefficients $\{b_i\}_{i=1}^p$. See Figure~\ref{fig:don_cnn} for an illustration.
Following \cite{hu&nan}, we construct the loss function using the full likelihood built upon the dataset $\{(x_i(\cdot),y_i)\}_{i=1}^n$, where $x_i(t)$ is observed at the grid points $\{t_1,t_2,...t_m\}$.  Specifically, the likelihood function has the following form:
\begin{eqnarray*}
    \label{eq:likelihood}
        L_n &=& \prod_{i=1}^n \lambda(y_i | x_i)\{1-F(y_i |x_i)\} \nonumber \\
        &=& \prod_{i=1}^n \exp\{h(x_i)(y_i)\}\exp\left\{-\int_{-\infty}^{y_i} e^{h(x_i)(s)}ds \right\}.
\end{eqnarray*}
Then the log likelihood becomes
\begin{equation} \label{eq:functional_loglikelihood}
        \ell_n = \sum_{i=1}^n \left[h(x_i)(y_i)-\int_{-\infty}^{y_i} e^{h(x_i)(s)}ds\right].
\end{equation}
We evaluate the above integrals using the Riemann summation. In particular, we first evaluate $h(x_i)(s)$ on an equal space partition, i.e. $s \in \{s_1, s_2, ..., s_k\}$, where $s_1$ is the first order statistic and $s_k$ is the maximum order statistic of $\{y_i\}_{i=1}^n$. Then, we obtain the following discretized loss function:
\begin{eqnarray}
        loss(h) = \frac{1}{n} \sum_{i=1}^n \sum_{j=2}^k I(s_{j-1} \le y_i)\left[ e^{h(x_i)(s_j)}(s_j-s_{j-1}) - h(x_i)(s_j)I(s_j \ge y_i)\right]. \label{eq:loss-functional}
\end{eqnarray}
Note that the above loss function is obtained by assigning a point mass of $1/n$ at $s_1$ for the case that the support of $Y$ has an unknown lower bound that may be $-\infty$.  Since $\widehat{F}(s_1 | x_i)=1/n$, we have $\int_{-\infty}^{s_1} e^{\widehat{h}(x_i)(s)}ds = -\log(1-1/n)$, here  $\widehat h$ denotes the estimator of $h$. Once $\widehat{h}$ is obtained, the conditional CDF given in (\ref{eq: cdf_functional}) can be estimated by
\begin{equation} \label{eq:functional_CDF-Estimator}
\widehat F(y|x) = I(s_1 \le y)\left\{1- (1-\frac{1}{n})\exp\left[ - \sum_{j=2}^k I(s_j \le y) e^{\widehat h(x)(s_j)} (s_j - s_{j-1}) \right] \right\}.
\end{equation}
From the above  (\ref{eq:functional_CDF-Estimator}) we see that $\widehat{F}(y  |  x) = 0$ when $y<s_1$ and $\widehat{F}(y  |  x) = 1/n$ when $y \in [s_1,s_2)$.
If the support of $Y$ has a known finite lower bound $s_0$, then the inner summation in (\ref{eq:loss-functional}) starts from $j=1$
and the conditional CDF estimator has the following form:
\begin{eqnarray*}
         \widehat F(y|x) = 1- \exp\left[ - \sum_{j=1}^k I(s_j \le y) e^{\widehat h(x)(s_j)} (s_j - s_{j-1}) \right].
\end{eqnarray*}
\section{Network Architecture and Hyperparameters}
We implement the proposed method directly in PyTorch~2.x. The branch net processes the functional covariate $x(t)$ with two blocks of 1D convolutional layers \citep{1dcnn}, each consisting of a Conv1D layer with ReLU activation followed by a max-pooling layer. The resulting feature vector is concatenated with the Euclidean covariates and passed through a two-layer fully connected network with ReLU activation. The trunk net is a fully connected network with two hidden layers. The hazard estimate $\widehat h(x)(y)$ is obtained as the inner product of the branch and trunk outputs \citep{deeponet}.
For the simulation and real data studies, the network architecture and the Adam \citep{adam} optimization settings are selected by Bayesian optimization with the Optuna library. Because the conditional distribution structures in our two simulation setups are very different (Gaussian with constant variance in Setup~1 vs.\ a heteroscedastic two-component exponential mixture in Setup~2), we run a separate Optuna study for each setup and report the configuration selected by each study. For Setup~1, the selected configuration uses two convolutional blocks of $32$ filters with kernel size $3$ and pool size $8$, a fully connected width of $256$, basis dimension $p=5$, an Adam learning rate of $2.9\times 10^{-3}$, and weight decay $2.6\times 10^{-5}$. For Setup~2, the selected configuration uses two convolutional blocks of $16$ filters with kernel size $5$ and pool size $8$, a fully connected width of $64$, basis dimension $p=20$, an Adam learning rate of $4.4\times 10^{-3}$, and weight decay $3.1\times 10^{-5}$. In both setups we train for at most $300$ epochs with early stopping on the validation loss. The $L_2$ baseline shares the same CNN feature extractor and uses a fully connected width of $64$, a standard regression-head size that keeps the baseline within the conventional capacity range while making the comparison with DeepONet fair. For the bike-sharing application, the selected configuration retains $32$ filters of kernel size $3$ but uses a smaller pool size of $2$, a fully connected width of $128$, basis dimension $p=16$, an Adam learning rate of $3\times 10^{-4}$, weight decay $1\times 10^{-4}$, and at most $400$ training epochs.
All neural networks are trained on a single NVIDIA A100 GPU.
\section{Simulations}
\label{sec:func_sim}
\begin{figure}[h!]
\centering
\includegraphics[width=0.6\linewidth]{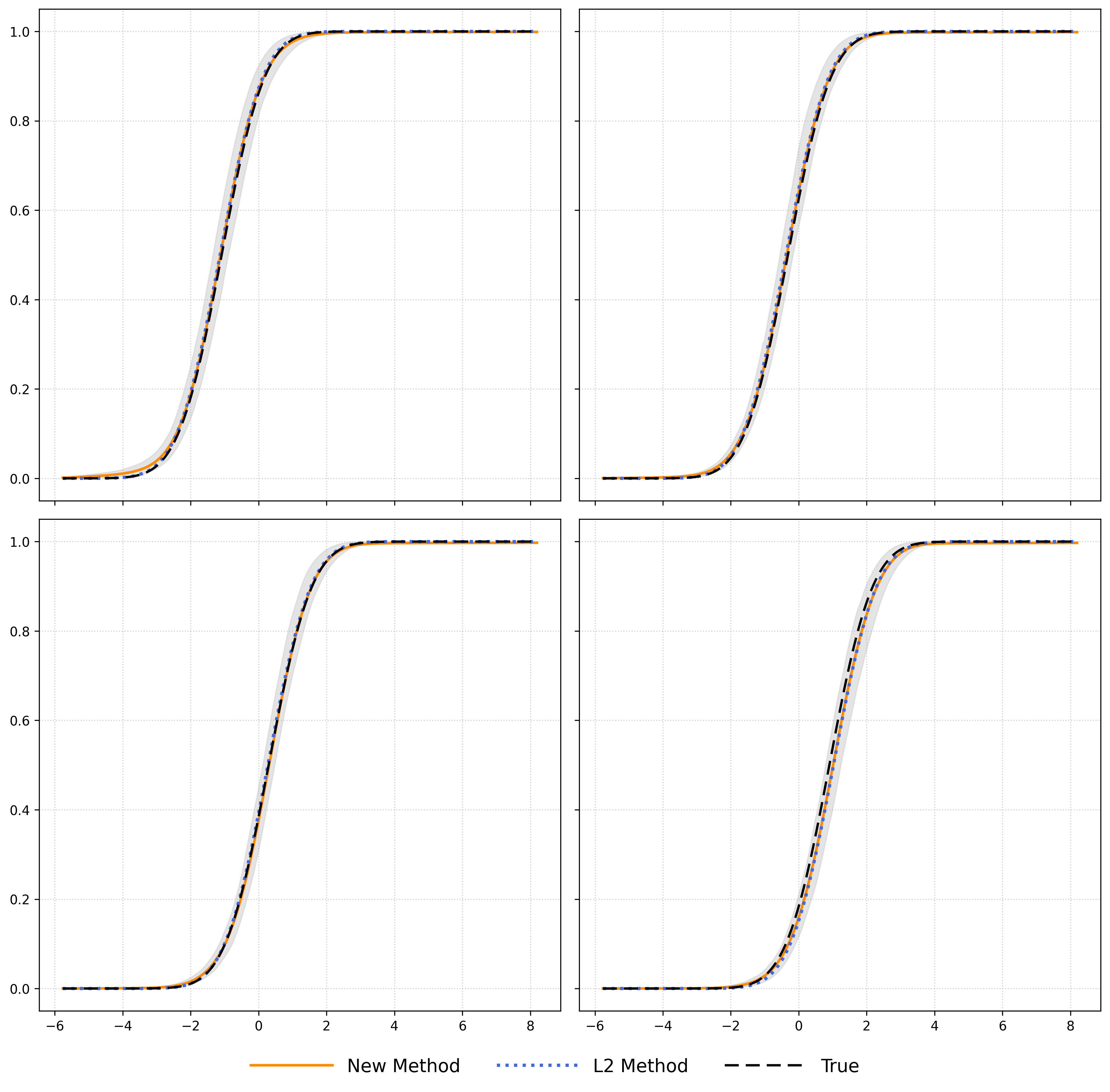}
\caption{Conditional distribution functions for $4$ representative sets of functional covariates (Setup 1). The gray shaded band is the $90\%$ empirical confidence interval of the new method across the $200$ replications.}
\label{fig:func_cnn_1}
\end{figure}

\begin{figure}[h!]
\centering
\includegraphics[width=0.6\linewidth]{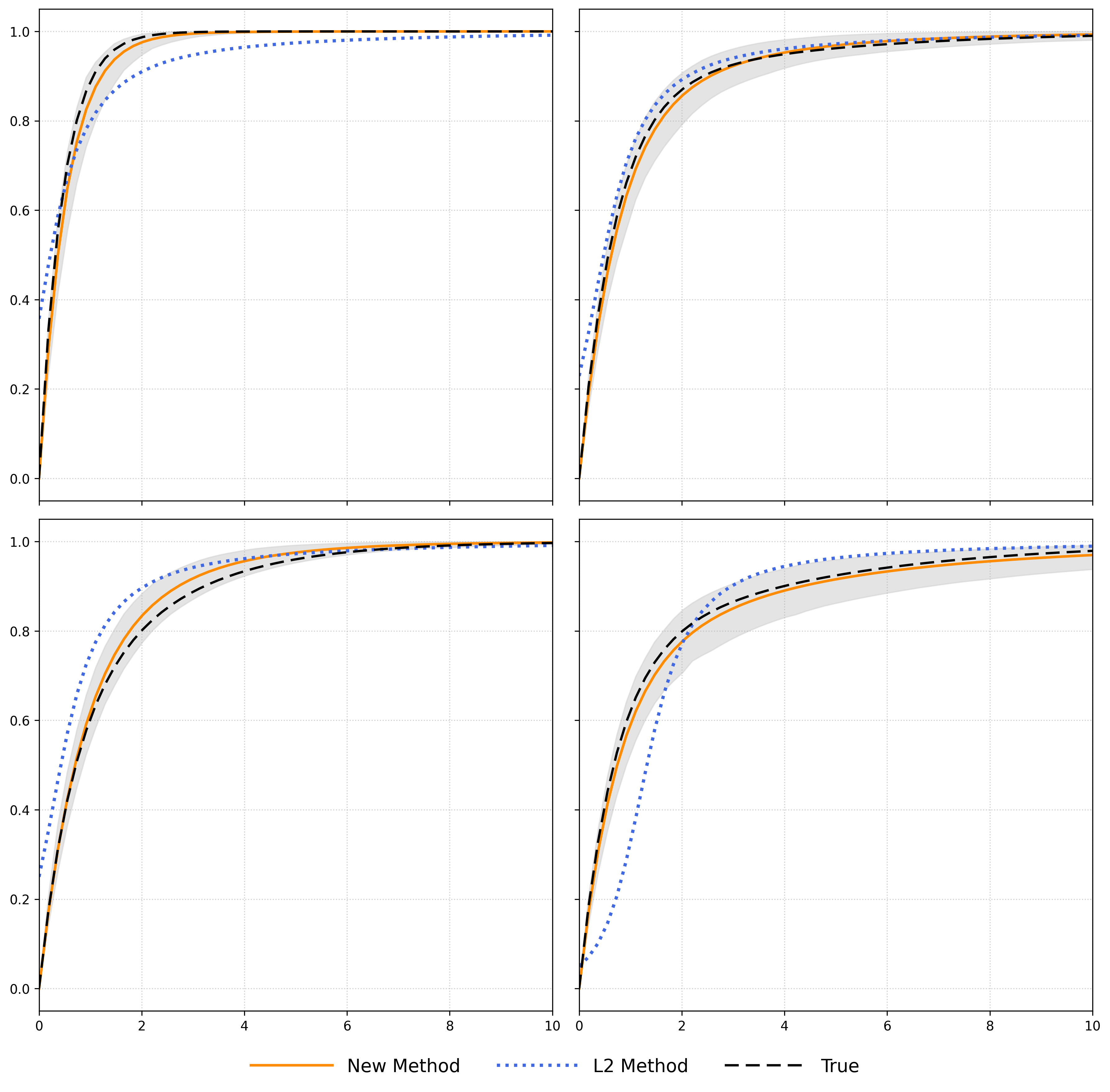}
\caption{Conditional distribution functions for $4$ representative sets of functional covariates (Setup 2). The gray shaded band is the $90\%$ empirical confidence interval of the new method across the $200$ replications.}
\label{fig:func_cnn_2}
\end{figure}

For $t \in \{0, \Delta t, 2\Delta t,....,1\}$ with $\Delta t = 0.01$, we generate $n$ independent copies of a functional covariate $x_{i}(t)$ from a standard Brownian motion $\{W_k\}_{k \in [0,1]}$ truncated to $[-3, 3]$, $i \in \{1,2,...n\}$. We generate another functional covariate $z_i(t)$ from a Poisson process with a rate of 10 truncated to $[0, 20]$. We also generate a time independent covariate $w_i$ from a standard normal distribution truncated to $[-3, 3]$.
We consider two simulation setups:
\begin{itemize}
    \item Setup 1: Generate response values from the following:
    \[
    y_i = \int_0^1 \{\sin[x_i(t)] + w_i\}[-\log (t)]\,dt + \epsilon_i,
\]
where $\epsilon_i \sim N(0,1)$ truncated to $[-3, 3]$.
\medskip
    \item {Setup 2:}
    Define
    \[
    \eta_i = 0.1 \int_0^1 z_i(t)\,dt + 0.1 \int_0^1 \sin[x_i(t)]\,dt + w_i,
    \qquad
    \pi_i = \frac{1}{1 + \exp(-\eta_i)}.
    \]
    Define two means with heterogeneous dependence on the functional covariates:
    \begin{eqnarray*}
    \mu_{i,1} &=& \exp\!\left(-0.5 + 0.02 \int_0^1 z_i(t)\,x_i(t)\,dt + 0.1\,w_i\right), \\
    \mu_{i,2} &=& \exp\!\left( \phantom{-}0.5 + 0.20 \int_0^1 z_i(t)\,x_i(t)\,dt + 0.6\,w_i\right).
    \end{eqnarray*}
    Then generate $y_i$ from a mixture of two exponential distributions
    \[
    y_i \sim \pi_i \,\text{Exponential}(1/\mu_{i,1}) + (1-\pi_i)\,\text{Exponential}(1/\mu_{i,2})
    \]
    truncated to $[0, 6\mu_{i,1}]$ and $[0, 6\mu_{i,2}]$, respectively. The two components have markedly different sensitivities to $\int_0^1 z_i(t)\,x_i(t)\,dt$ and $w_i$, so the conditional variance of $Y$ varies strongly with the covariates, violating the constant-error-variance assumption underlying the $L_2$ method.
\end{itemize}
Note that we truncate those randomly generated quantities to meet the sufficient conditions of the Universal Approximation Theorem for operators \citep{operator_universal}.
We independently generate training sets and validation sets. For each simulation run, when the validation loss no longer decreases, we stop training to avoid overfitting. Once the neural net model is trained, we use the fitted model to estimate the conditional CDF curves given newly generated covariates. We repeat the process for $N=200$ independent replications, then plot the sample average and 90\% empirical confidence band of the estimated conditional CDF curves for the same set of covariates.
The traditional neural network method with the commonly used $L_2$ loss function gives the conditional mean estimator. The conditional distribution function given a set of covariate values in such a case is commonly estimated by shifting the center of the empirical distribution of training set residuals to the estimated conditional mean. This would yield a valid estimator under the assumption that the errors (outcomes subtract their conditional means) are i.i.d. and uncorrelated with conditional means. On the other hand, in our method, an estimator of the conditional distribution function gives a conditional mean estimator as follows:
\begin{equation} \label{ConditionalMean}
    \int_{-\infty}^{\infty}td\widehat F(t|x) = \sum_{i=1}^n t_i \left(\widehat F(t_i|x) - \widehat F_k(t_{i-1}|x)\right).
\end{equation}
Thus we can compare our method to the widely imposed mean regression method with the same CNN neural network structure on the estimation of the conditional distribution function as well as the estimation of the conditional mean.
Conditional CDF curves for $4$ representative sets of functional covariates, are shown in Figures~\ref{fig:func_cnn_1} and \ref{fig:func_cnn_2} for setups 1 and 2, respectively. We can see that, in Setup 1, both methods estimate CDF well.
In Setup 2, because the two mixture components have very different sensitivities to the functional and scalar covariates, the conditional variance of $Y$ is itself a function of the covariates. Therefore the $L_2$ method, which shifts a single pooled empirical residual distribution to the predicted conditional mean, cannot recover the correct shape of the conditional CDF, while the proposed method yields conditional CDF estimates that follow the true conditional CDFs closely.
To quantify these comparisons, in each replication we evaluate both methods on an independent test set of $n_{\mathrm{test}} = 500$ observations. Given an estimated conditional CDF $\widehat F(\cdot \mid x)$, a commonly used level-$(1-\alpha)$ prediction interval for a data point with covariate $x$ is taken to be the equal-probability-tailed interval
\[
[\ell(x;\alpha),\, u(x;\alpha)] \;=\; \big[\widehat F^{-1}(\alpha/2 \mid x),\; \widehat F^{-1}(1-\alpha/2 \mid x)\big],
\]
where
\[
\widehat F^{-1}(q \mid x) = \inf\{s: \widehat F(s \mid x) \ge q\}.
\]
For the proposed method, $\widehat F$ is the estimator in (\ref{eq:functional_CDF-Estimator}). For the $L_2$ method, the residual-shift estimator described above is $\widehat F_{L_2}(s \mid x) = \widehat F_r\{s - \widehat \mu(x)\}$, where $\widehat \mu$ is the fitted mean and $\widehat F_r$ is the empirical distribution of the training residuals $r_i = y_i - \widehat \mu(x_i)$.
We compare these two methods with the estimation precisions of $\widehat\mu$ and $\widehat F$, and both marginal and conditional coverage probabilities of their prediction intervals. Table~\ref{tab:func_mse} summarizes these comparisons from simulations with $200$ replications.
Marginal coverage rates in Table~\ref{tab:func_mse} are the proportions of test data whose responses fall within their prediction intervals, averaged over $200$ independent replications. Conditional coverage rates, on the other hand, are first evaluated in a set of bins determined by $\widehat\mu$ and then compared to the target probability, which is summarized using mean squared errors (MSE). Specifically, within each replication we sort the test data by their predicted conditional means $\widehat \mu(X_i)$ that are obtained from either the network output for the $L_2$ method or Equation (\ref{ConditionalMean}) for the proposed method. Then we partition the sorted means into $B = 10$ consecutive bins $\mathcal{I}_1, \ldots, \mathcal{I}_B$ of equal size, so that data in a bin have similar predicted locations. In bin $b$, the empirical coverage probability is
\[
  \widehat{\mathrm{CP}}_b(\alpha)
   \;=\; \frac{1}{|\mathcal{I}_b|}
   \sum_{i \in \mathcal{I}_b} \mathbf{1}\!\big\{\, y_i \in [\ell(x_i;\alpha),\, u(x_i;\alpha)] \,\big\},
\]
which should be close to the target $1-\alpha$ for every bin if the conditional distribution is well estimated. The conditional coverage probability MSE aggregates the squared deviations across bins, weighted by bin size:
\[
\mathrm{MSE}_{\mathrm{CCP}}(\alpha)
\;=\; \sum_{b=1}^B
\frac{|\mathcal{I}_b|}{n_{\mathrm{test}}}
\Big(\widehat{\mathrm{CP}}_b(\alpha) - (1-\alpha)\Big)^2,
\qquad
n_{\mathrm{test}} = \sum_{b=1}^B |\mathcal{I}_b|,
\]
where the weights reduce to $1/B$ for equal-sized bins. The reported $\mathrm{MSE}_{\mathrm{CCP}}$ values are averaged over the $200$ replications.
From Table~\ref{tab:func_mse} we see similar MSEs for $\widehat\mu$ obtained from both methods. The $L_2$ method yields a slightly smaller MSE for $\widehat F$ in Setup 1 where the error is independent of mean, but a much larger MSE in Setup 2 where the error distribution depends on covariates. This is expected because the $L_2$ method is optimal in Setup 1.
The marginal coverage rates in Table~\ref{tab:func_mse} are close to the nominal values for both methods in both setups. This is not surprising for both methods in Setup 1 and the new method in Setup 2. The good result of the $L_2$ method in Setup 2, although the error distribution depends on covariates, can be explained in the following. The coverage probability of the $L_2$ method is determined by
\begin{eqnarray*}
P\left\{ Y - \widehat \mu(X) \in \left[{\widehat F}^{-1}_{r,\alpha/2},\, {\widehat F}^{-1}_{r,1-\alpha/2}\right] \right\}.
\end{eqnarray*}
When the sample size is large and $\widehat \mu$ converges to $\mu$ in probability, it would be expected that $\widehat F_r$ can approximate well the distribution of $Y-\mu(X)$, so the above probability approaches $1-\alpha$ for independent random variables $(X,Y)$.
The largest differences between the two methods are observed in conditional coverage in Setup 2. We see from Table~\ref{tab:func_mse} that the two methods achieve similar MSEs for conditional coverage in Setup 1, where the error distribution is independent of covariate. But in Setup 2, the MSE of the $L_2$ method is about 10 times worse, indicating the desirable robustness and superiority of our proposed new method.
\begin{table}[h!]
    \centering
    \begin{tabular}{l r r r r}
         \hline
         & \multicolumn{2}{c}{Setup 1} & \multicolumn{2}{c}{Setup 2} \\
         \cline{2-3} \cline{4-5}
          & \multicolumn{1}{c}{New Method} & \multicolumn{1}{c}{L2 Method}
                & \multicolumn{1}{c}{New Method} & \multicolumn{1}{c}{L2 Method} \\
         \hline
         MSE for $\widehat{\mu}$                  &  1.02 &  1.01 &  6.34 &   6.37 \\
         MSE for $\widehat{F}$ ($\times 10^{-4}$) &  5.74 &  2.36 &  2.83 &  16.75 \\
         90\% Coverage                            & 89.93 & 89.45 & 89.07 &  89.95 \\
         95\% Coverage                            & 95.17 & 94.66 & 94.37 &  94.96 \\
         $\mathrm{MSE}_{\mathrm{CCP}}(0.1)$ ($\times 10^{-4}$)  & 20.81 & 19.38 & 21.39 & 258.95 \\
         $\mathrm{MSE}_{\mathrm{CCP}}(0.05)$ ($\times 10^{-4}$) & 10.02 & 10.66 & 11.45 & 106.22 \\
         \hline
    \end{tabular}
    \caption{Average MSE for mean estimation ($\widehat{\mu}$), MSE for conditional distribution estimation ($\widehat{F}$), prediction coverage rates, and conditional coverage probability MSE $\mathrm{MSE}_{\mathrm{CCP}}(\alpha)$ at nominal coverage $1-\alpha$, under Setup~1 and Setup~2, averaged over 200 replications. Hyperparameters for the proposed method are selected by Optuna.}
    \label{tab:func_mse}
\end{table}
\section{Application to Bike Sharing data analysis}
\label{sec:func_real}
In this section, we apply our method to a real world data set discussed in \cite{dlfunctionalinput}. The Bike Sharing data \citep{misc_bike_sharing_275} contains the hourly and daily count of rental bikes between years 2011 and 2012 in Capital bike share system with the corresponding weather and seasonal information. The goal is to predict the number of daily rentals given hourly temperature (functional, 24 points) in each day. We also include a binary variable representing working day (yes/no) in the prediction model because this could be a crucial factor in bike rental. The sample size is 730. Although the number of daily rentals is discrete, it ranges from a few hundreds to more than 8000. So, it would be reasonable to approximate the response as a continuous variable.
The Bike dataset has been previously analyzed using other FDA methods, such as in \cite{dlfunctionalinput}, but the data usage differs, and the preprocessing details were not clearly presented in earlier works. For example, \cite{dlfunctionalinput} only uses Saturday records to eliminate the effect of different days of the week, resulting in a very small sample size. These factors make it challenging to directly compare different methods with existing results.
In \cite{dlfunctionalinput}, it was demonstrated that conventional mean regression neural networks with a CNN structure perform well in predicting mean outcomes compared to kernel methods. Therefore, we compare our method, which predicts the conditional CDF, with the mean regression neural networks ($L_2$ method). Specifically, we employ the same CNN structure in the branch net of our method as used in the conventional mean regression neural networks.
We use 5-fold cross-validation to train the models and make predictions. In real-world datasets, where the underlying conditional CDF is unknown, we assess the quality of the CDF estimate by examining the coverage rate. Additionally, as described in Section~\ref{sec:meth}, we calculate the conditional mean from the estimated conditional CDF. The performance of the methods is then evaluated using 5-fold cross-validated MSE
for mean estimation. The results are summarized in Table~\ref{tab:performance_comparison}, which illustrates much improved conditional coverages of our proposed method.
\begin{table}[h!]
    \centering
    \begin{tabular}{l r r}
         \hline
          & \multicolumn{1}{c}{New Method} & \multicolumn{1}{c}{L2 Method} \\
         \hline
         MSE for $\widehat{\mu}$                  &  0.48 &   0.49 \\
         90\% Coverage                            & 87.41 &  78.52 \\
         95\% Coverage                            & 93.16 &  86.32 \\
         $\mathrm{MSE}_{\mathrm{CCP}}(0.1)$ (\%)  & 95.92 & 272.15 \\
         $\mathrm{MSE}_{\mathrm{CCP}}(0.05)$ (\%) & 54.51 & 161.90 \\
         \hline
    \end{tabular}
    \caption{MSE for mean estimation ($\widehat{\mu}$), prediction coverage rates, and conditional coverage probability $\mathrm{MSE}_{\mathrm{CCP}}(\alpha)$ at nominal $1-\alpha$ coverage, for the bike-sharing data, evaluated by 5-fold cross-validation.}
    \label{tab:performance_comparison}
\end{table}
\section{Discussion}
It is worth noting that directly using the full likelihood enables our method to be fully non-parametric, providing maximum robustness with minimal number of assumptions. This makes our approach always valid across a broad range of data distributions, ensuring reliability in various situations. Other methods may be more efficient, but only when their specific model assumptions hold.
Our method is different from the well-known generative deep learning models. Generative models, such as Generative Adversarial Networks (GANs) \citep{gan}, Variational Autoencoders (VAEs) \citep{vae}, and Diffusion Models \citep{diffusion}, provide ways to draw new samples from the empirical data distribution,
whereas our method directly estimates an arbitrary conditional distribution function. It would be interesting future work to compare our method with those generative deep learning methods within the context of FDA.
\section*{Disclosure statement}
No potential conflict of interest was reported by the author(s).
\section*{Declaration of generative AI use}
The authors report generative AI was not used in their research or preparation of this manuscript. 
\section*{Funding}
This work was supported by the NIH under Grant RF1 AG075107; and NSF under Grant DMS 2412746.
\section*{Data availability statement}
The Bike Sharing data that support the findings of this study are openly available in the UCI Machine Learning Repository at https://doi.org/10.24432/C5W894 \citep{misc_bike_sharing_275}.
\bibliographystyle{tfcad}
\bibliography{interactcadsample}
\end{document}